\documentclass{article}
\usepackage[utf8]{inputenc}
\usepackage[english]{babel}
\usepackage[T1]{fontenc}
\usepackage{amssymb, graphicx, amsmath}
\usepackage{algpseudocode}
\usepackage{algorithm}
\DeclareMathOperator*{\argmin}{argmin}
\DeclareMathOperator*{\argmax}{argmax}

\title{Module-based regularization improves Gaussian graphical models when observing noisy data}
\author{Magnus Neuman$^1$, Joaquín Calatayud$^2$, Viktor Tasselius$^{1,3}$,  Martin Rosvall$^1$}
\date{}

\begin{document}

\maketitle
\begin{small}
    \noindent 1. Integrated Science Lab, Department of Physics, Umeå University, Umeå, Sweden\\
    \noindent 2. Department of Biology, Geology, Physics and Inorganic Chemistry, King Juan Carlos University, Madrid, Spain\\
    \noindent 3. School of Public Health and Community Medicine, University of Gothenburg, Gothenburg, Sweden
    
\end{small}

\section*{Abstract}
Inferring relations from correlational data allows researchers across the sciences to uncover complex connections between variables for insights into the underlying mechanisms. The researchers often represent inferred relations using Gaussian graphical models, requiring regularization to sparsify the models. Acknowledging that the modular structure of the inferred network is often studied, we suggest module-based regularization to balance under- and overfitting. Compared with the graphical lasso, a standard approach using the Gaussian log-likelihood for estimating the regularization strength, this approach better recovers and infers modular structure in noisy synthetic and real data. The module-based regularization technique improves the usefulness of Gaussian graphical models in the many applications where they are employed.


\section*{Introduction}
Inferring relations between observed features from correlational data is a foundational approach to exploring underlying mechanisms in, for example,  ecological, genetic and neural systems \cite{Barberan, Wang, Bullmore}. The resulting relations are often represented as a network where the features are nodes and their respective relations are links. These networks are dense, making it difficult to discern relevant structures. Field-specific methods to sparsify them suggest soft thresholding \cite{Horvath}, but hard thresholding is often applied in practice \cite{Barberan, deVries, Neuman}. 
Gaussian graphical models provide an alternative way of representing correlational data by encoding relations between features through partial correlations. 
A popular approach to infer a Gaussian graphical model is the graphical lasso (GLASSO) \cite{Friedman2007, Yuan}, which estimates the precision matrix while ensuring sparsity through $l_1$-regularization. This method and related methods, such as neighborhood selection \cite{Meinshausen}, elastic net \cite{Zou_elasticnet} and Markov networks \cite{Murphy}, are widely used in many disciplines \cite{Harris, Epskamp_psycho, Cao_cElegans,Severson_batteries}. Despite its widespread application, GLASSO struggles to tackle noise and the high dimensionality that comes with many observed features, often exceeding the number of available samples \cite{Raskutti2008, Wainwright2009, Ravikumar2011, Liu2012}. 

Representing the inferred relations as networks enables studying structures in the data with standard tools from network science. Network modules -- groups of tightly connected nodes -- are studied across scientific disciplines because they reveal significant patterns and functional relationships in diverse systems, ranging from ecological \cite{calatayud2020positive} to metabolic networks \cite{guimera2005functional}. However, the GLASSO is agnostic to modular structure in the inferred networks, which can obscure network structure and subsequent interpretation and understanding of the studied systems. Simultaneously inferring the network and its modular structure can alleviate this problem, but requires prior knowledge about the dynamical processes on the network \cite{Peixoto}. Attempts at integrating modular structure with the GLASSO use a varying penalization term that depends on the underlying modular structure, with a predetermined number of modules \cite{Ambroise2009, Steeg2019}. This approach limits the usability since the number of modules is in general unknown. Manually setting the number of modules risks over- or underfitting the modular structure, and nodes in data with high noise rarely have a significant module membership.

Here we integrate the two steps from relational data to network modules -- network inference and community detection -- in an extension of the GLASSO method. We use the network's modular structure to select the regularization strength, which allows us to balance over- and underfitting the modular structure to the data. Using synthetic data, we show that this approach allows us to recover more modular structures in noisy data compared with the standard GLASSO. Applied to country-level daily incidence during the Covid-19 pandemic and gene co-expression data from the plant {\it Arabidopsis thaliana}, we find that the module-based GLASSO can identify more modular structure in these data compared to the standard GLASSO -- highly relevant for researchers studying these systems. 

\section*{Results}
Gaussian graphical models describe relations between observed features. They are derived from the precision matrix $\Theta$ that encodes conditional independence between variables, meaning that two observations $X_i$ and $X_j$ are independent, given all other observations, if the corresponding $ij$:th element in $\Theta$ is zero. The GLASSO aims at maximizing the Gaussian log-likelihood of the precision matrix given the data while ensuring a sparse solution by imposing an $l_1$-regularization term $\lambda ||\Theta ||_1$, with the regularization parameter $\lambda$. The best precision matrix $\Theta^{\lambda}$ for a specific value of $\lambda$ is thus 
\begin{equation}
    \Theta^{\lambda} = \argmax_{\Theta}\left( \log \det (\Theta) -\mathrm{tr} (\Theta \hat{\Sigma}) - \lambda ||\Theta ||_1 \right),
\end{equation}
where $\hat{\Sigma}$ is the covariance matrix calculated from the observed data. The parameter $\lambda$ determines the regularization strength and thereby the sparsity of the inferred precision matrix. The regularization parameter $\lambda$ is often determined using cross-validation where the best value $\lambda^*$ is the one that has the largest log-likelihood of the test data $\hat{\Sigma}^{test}$ given the model $\Theta^{\lambda, train}$ inferred from the training data $\hat{\Sigma}^{train}$ such that
\begin{equation}
    \lambda^* = \argmax_{\lambda}\left( \log \det (\Theta^{\lambda, train}) -\mathrm{tr} (\Theta^{\lambda, train} \hat{\Sigma}^{test})\right).
\label{eq:loglik}
\end{equation}
The resulting regularization strength conserves relations with support in both the training and test data, without considering any conserved structures in the data.

To take the modular structure into account when selecting the regularization strength, we suggest using the map equation framework and its search algorithm Infomap \cite{RosvallPNAS2008, Rosvall2, Edler1}. The map equation encodes a random walk on a network and measures the codelength $L(M)$ of the random walk given a partition $M$ of the network into modules. Infomap uses a greedy approach to find the partition $M^*$ that minimizes the codelength, 
\begin{equation}
    M^* = \argmin_M L(M),
\end{equation}
such that $M^*$ is the best partition of the network according to the minimum description length principle. This popular approach is widely recognized as one of the best methods for detecting network communities \cite{Lancichinetti,Aldecoa2013}. To connect Infomap with the GLASSO regularization, we suggest maximizing the signal of modular structure present in both the training and test sets when cross-validating the regularization parameter $\lambda$. We measure this signal using the codelength savings in the test data given the optimal partition of the training data, such that
\begin{equation}
    \lambda^* = \argmax_{\lambda}  \frac{L^{test}(1)-L^{test}(M^{\lambda, train})}{L^{test}(1)},
\label{eq:clsav}
\end{equation}
where $M^{\lambda, train}$ is the optimal partition of the training data and $L^{test}(1)$ is the one-level codelength of the test data with all nodes in the same module. The 
codelength savings are positive if the modular structure in the training data is present also in the test data and has its maximum when this shared modular structure is most prominent. This peak is associated with the $\lambda$ that best captures modular structure in the data without over- or underfitting, analogous to the log-likelihood in Eq.~\ref{eq:loglik}. In previous work \cite{Neuman}, we explored this module-based approach for hard thresholding of correlation networks and showed that a too low threshold gives a highly connected network with little modular structure in both the training and test networks, while a too high threshold gives a highly modular structure in the training network that is not present in the test network. The same reasoning applies to the GLASSO when selecting regularization strength. The approach we suggest finds the best compromise between these two extremes.

To derive the network $\mathcal{G}(\Theta^*)$ that balances under- and overfitting, we use GLASSO to estimate the precision matrix $\Theta^*$  corresponding to $\lambda^*$. We use the relation between a partition matrix element and the partial correlation so that the link $e_{ij}$ between nodes $i$ and $j$ is given by 
\begin{equation}
    e_{ij} = |- \theta_{ij}/ \sqrt{\theta_{ii}\theta_{jj}}|,
\end{equation}
where $\theta_{ij}$ is elements of $\Theta^*$, and the link weight is thus the absolute value of the partial correlation.

\subsection*{Synthetic data}
To test the module-based regularization, we generate synthetic data by sampling a covariance matrix $S$ from a Wishart distribution such that
\begin{equation}
    S\sim W_p(n, \Sigma),
\end{equation}
where $\Sigma$ is the block-diagonal covariance matrix of the planted (oracle) modular structure, $p$ is the dimension (number of features or nodes) and $n$ is the number of degrees of freedom. We plant a modular structure by imposing a block-diagonal structure:
\begin{equation}
    \Sigma_{i,j} = 
    \begin{cases}
        1, & i=j \\
        c, & M(p_i)=M(p_j)\\
        0, & M(p_i)\neq M(p_j),
    \end{cases}
\end{equation}
where $M(p_i)$ denotes the module of node $p_i$. In this way, both the planted covariance matrix $\Sigma$ and the sampled matrix $S$ are positive definite. To change the signal-to-noise ratio, we can vary the the planted within-module covariance $c$ and the degrees of freedom $n$ in the Wishart distribution. Using this setup, we sample the observed data $X$ from a $p$-variate normal distribution such that $X_i\sim N_p(0, S)$ and $X\in \mathbb{R}^{p\times q}$, where $q$ denotes the number of samples. The objective is to infer the planted modular structure using these data.

\begin{figure}[tb]
\centering
\includegraphics[scale=0.6]{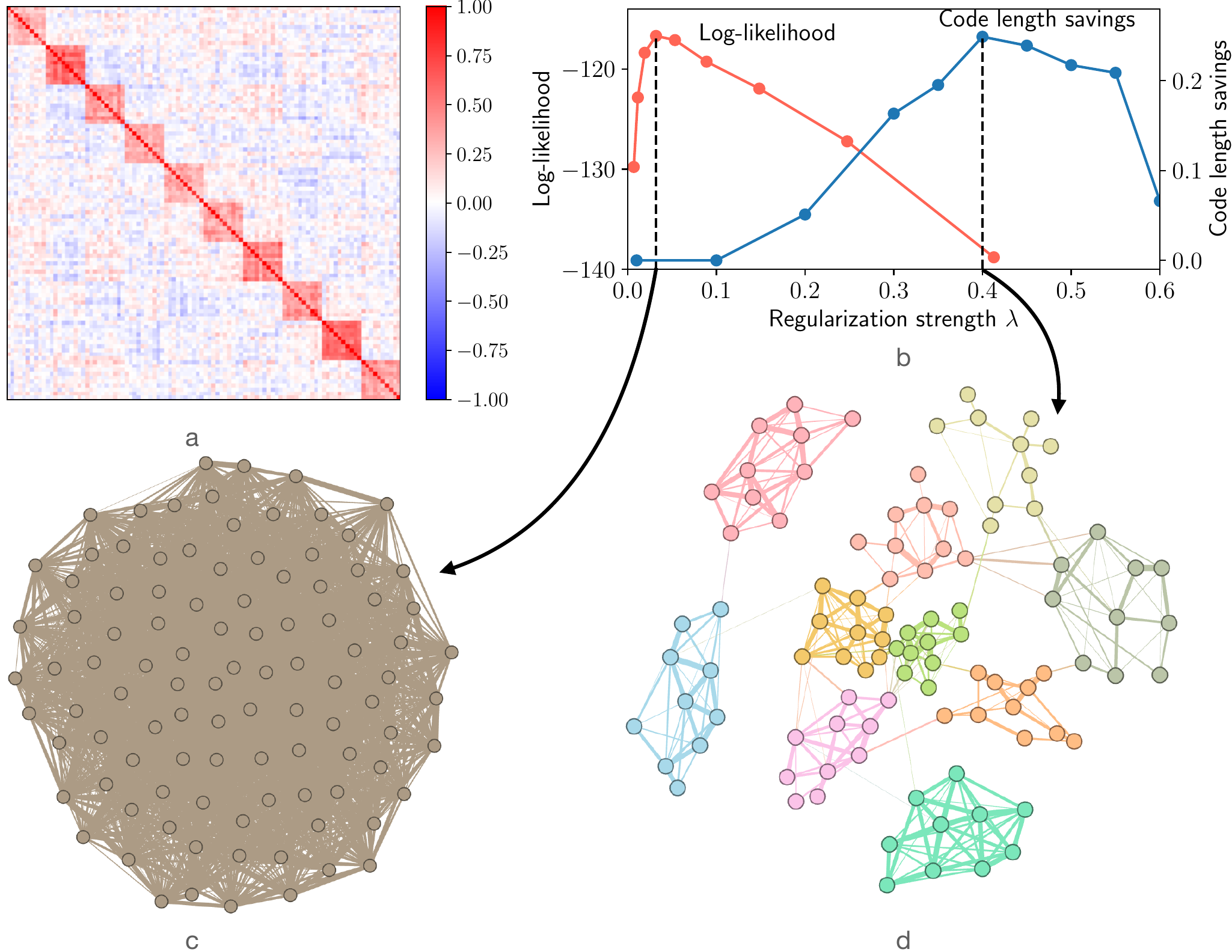}
\caption{\textbf{Comparing Standard and Modular GLASSO methods in detecting planted modular structure}. The covariance matrix sampled from the Wishart distribution is noisy but with modular structure (a). With data sampled using this matrix, the GLASSO based on log-likelihood (Standard GLASSO) regularizes less than the GLASSO based on modular structure through Infomap's codelength (Modular GLASSO) (b), which leads to the Standard GLASSO's failure to identify any modular structure (c) while the Modular GLASSO successfully recovers the planted modular structure (d).}   
\label{fig:1x4}
\end{figure}

We use ten planted modules with ten nodes in each module to illustrate our approach, as shown in Fig.~\ref{fig:1x4}. The sampled covariance matrix is shown in Fig.~\ref{fig:1x4}a, where the number of degrees of freedom is $n=100$ and the planted covariance is $c=0.4$, and we see that the matrix is noisy but with discernible modular structure. Using this covariance matrix we draw $q=100$ samples to obtain the synthetic data. We see that the log-likelihood-based GLASSO (hereafter Standard GLASSO) gives a lower optimal $\lambda$ value and hence regularizes less than the module-based GLASSO (hereafter Modular GLASSO), since their respective quality functions peak at different $\lambda$-values (Fig.~\ref{fig:1x4}b). This leads to the Standard GLASSO including a lot of noisy links, as the network representation shows (Fig.~\ref{fig:1x4}c). In contrast, Modular GLASSO increases the regularization to maximize the modular structure common to the test and training data, enabling the method to correctly recover the planted modular structure (Fig.~\ref{fig:1x4}d).


\begin{figure}[tb]
\centering
\includegraphics[scale=0.8]{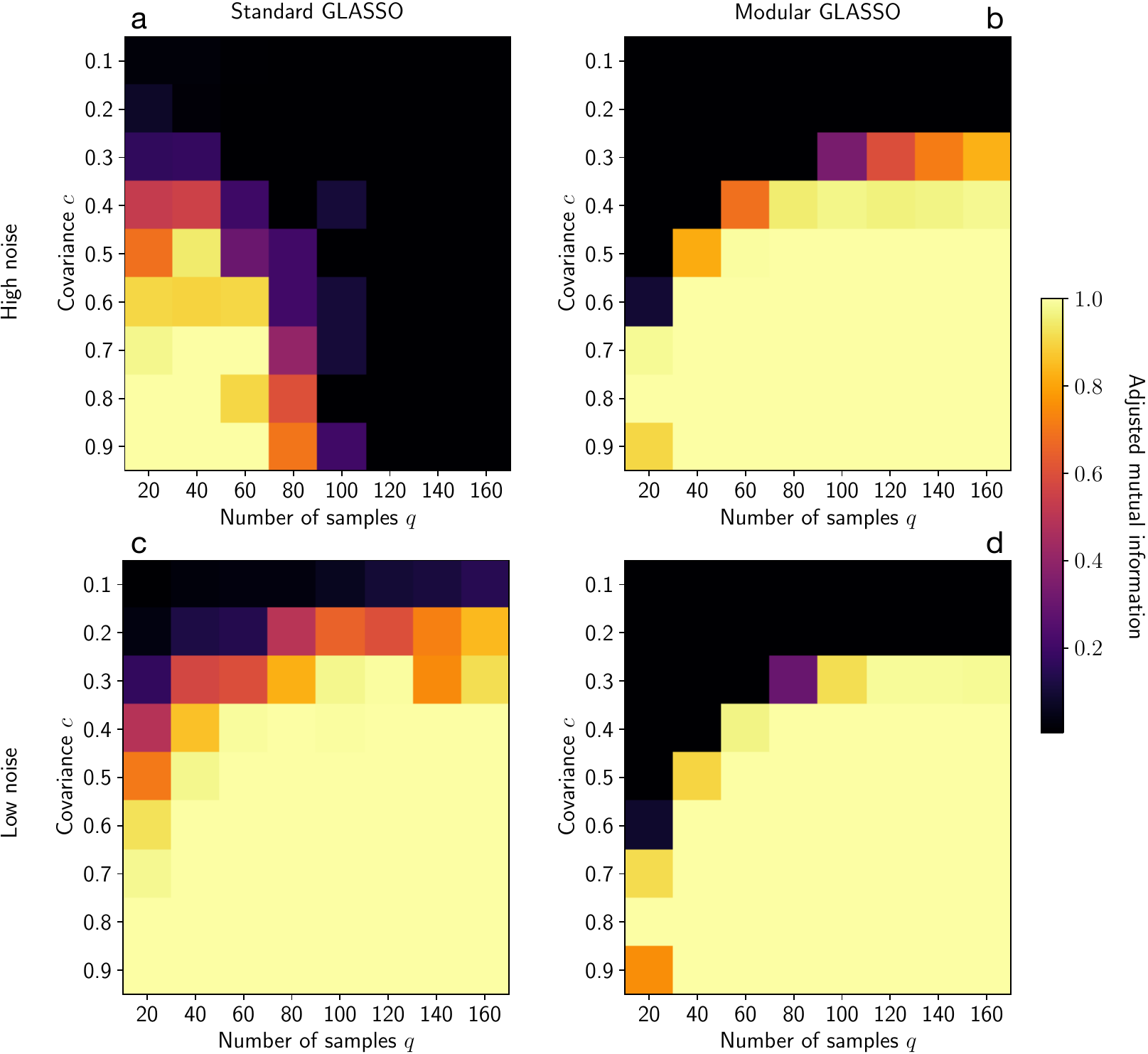}
\caption{\textbf{Performance comparison of the Standard and the Modular GLASSO in detecting planted partitions under low and high noise conditions}. The adjusted mutual information (AMI) between recovered and planted partitions shows that the Standard GLASSO finds the planted partition only if the samples are few when the noise level is high, but when samples and within-module covariance are sufficient for low noise. In contrast, the Modular GLASSO finds the planted partition also in high noise when samples and covariance are sufficient.}
\label{fig:2}
\end{figure}

To explore this result, we vary the covariance $c$ and the number of samples $q$. We quantify how well the methods recover the planted partition by calculating the adjusted mutual information (AMI) between the planted partition and the recovered partition, which is the partition found by Infomap given the network $\mathcal{G}(\Theta^*)$ (Fig.~\ref{fig:2}). The Standard GLASSO recovers the planted partition when the number of samples is small and the covariance is large, but not for many samples (Fig.~\ref{fig:2}a). This tendency to recover more modular structure with fewer samples exemplifies the ``blessing of dimensionality'' \cite{Steeg2019}. In contrast, the Modular GLASSO recovers the modular structure when the number of samples and the covariance are large -- increasing the number of samples is always beneficial until all modular structure is recovered (Fig.~\ref{fig:2}b).

When we decrease the noise level by using $n=1000$ degrees of freedom in the Wishart distribution, the methods show similar performance, with a slight advantage for the Standard GLASSO, and recover the modular structure for sufficiently large covariance and number of samples (Fig.~\ref{fig:2}cd). This result indicates that Standard GLASSO's performance is sensitive to the presence of noise in the data. 

\begin{figure}[tb]
\centering
\includegraphics[scale=0.8]{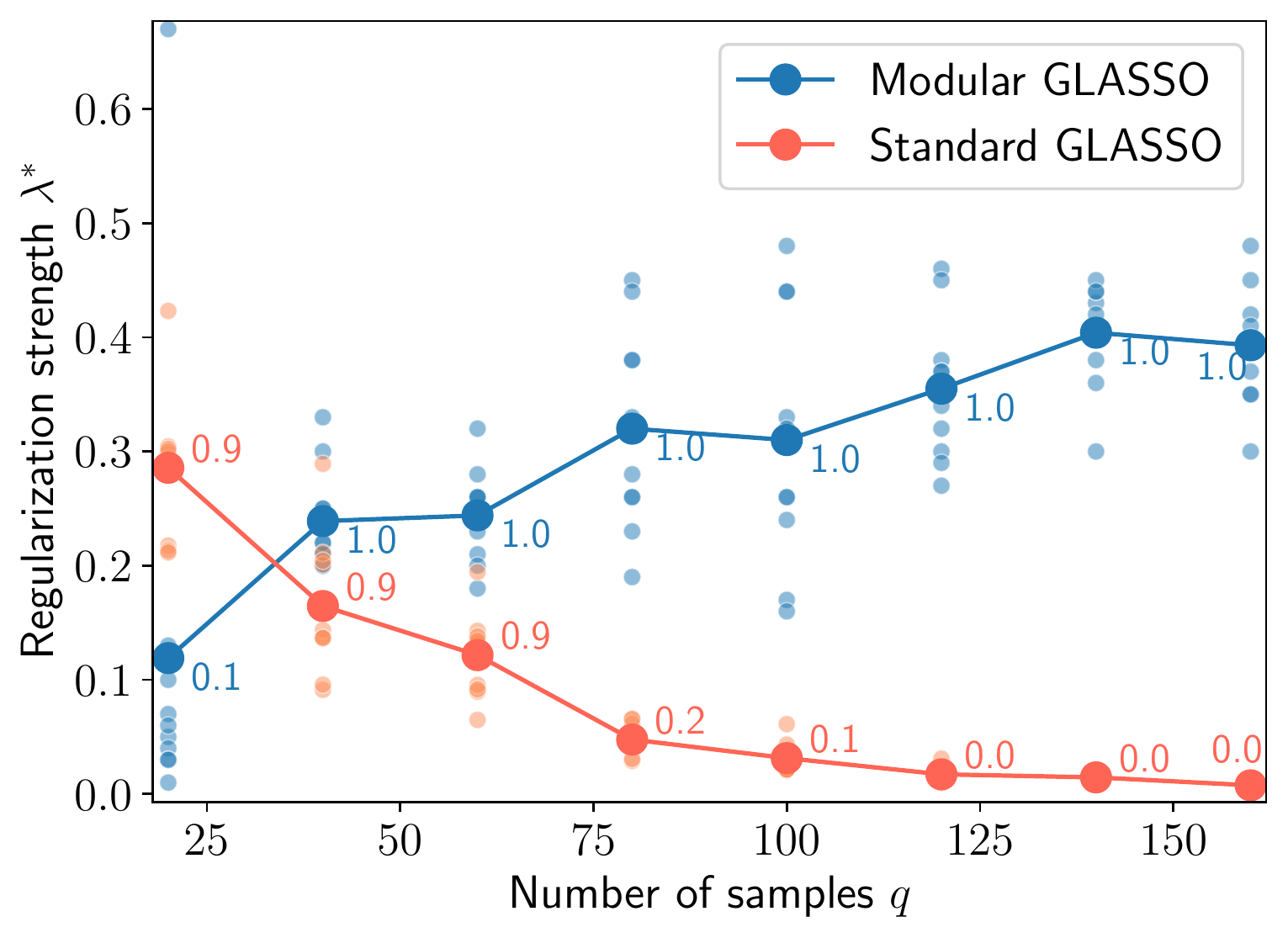}
\caption{\textbf{The optimal regularization strength as a function of the number of samples for Standard and Modular GLASSO}. The Standard GLASSO regularizes less, resulting in the inclusion of many noisy correlations. The Modular GLASSO regularizes based on the modular structure, leading to a stronger regularization as the number of samples increases and the recovery of the planted modular structure. The large points represent averages over ten runs, with individual runs shown as small points. The AMI between recovered and planted partitions is displayed as a number next to each point.}   
\label{fig:3}
\end{figure}

To compare the methods more closely, we plot the optimal $\lambda$-value as a function of the number of samples for a fixed value $c=0.6$ of the within-module covariance (Fig.~\ref{fig:3}). The Standard GLASSO's optimal $\lambda$ decreases for more samples, while it increases for the Modular GLASSO. The AMI approaches zero for the Standard GLASSO for more samples because it regularizes less. The Modular GLASSO's regularization increases with the number of samples. The AMI reaches 1 since the method captures the signal of the modular structure and adapts the regularization. In contrast, the Standard GLASSO does not regularize at all when there are many samples, retaining all spurious relations between the observed features and obscuring the modular structure.


 \subsection*{Real-world data}
\begin{figure}[tb]
\centering
\includegraphics[scale=0.48]{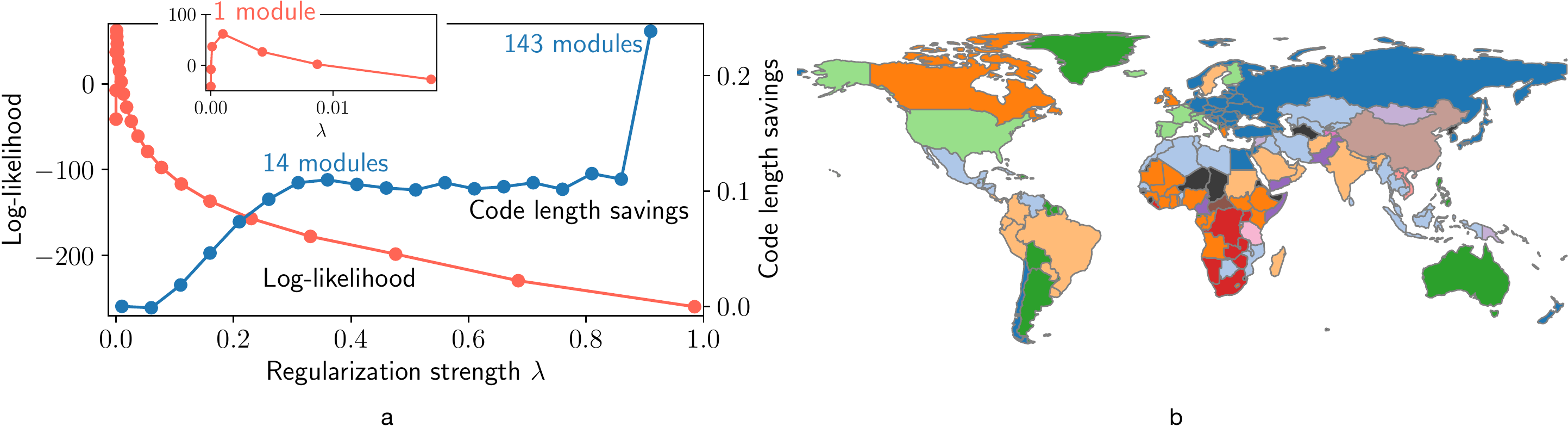}
\caption{\textbf{Application of Standard and Modular GLASSO to Covid-19 incidence data}. The Standard GLASSO (log-likelihood) and Modular GLASSO (codelength savings) suggest vastly different regularization strengths for the Covid-19 data (a). The Standard GLASSO reveals no modular structure in the resulting network, while the Modular GLASSO uncovers the 14 modules represented by different colors on the world map (b). The modules exhibit a geographical signal as adjacent countries tend to belong to the same module, with some interesting exceptions.}
\label{fig:4}
\end{figure}

\paragraph{Covid-19 data.} We analyze the global Covid-19 data \cite{covid} with the daily incidence of Covid-19 in 192 countries over 777 days, from 2020/01/01 to 2022/02/15. The observed features are the world's countries and the samples are the 777 days with Covid-19 incidence, making it a sample-rich data set. The signal-to-noise ratio is high because the distribution of correlations significantly deviates from what would be expected from spurious correlations (the Kolmogorov-Smirnov statistic is 0.72). For these data, Standard and Modular GLASSO suggest vastly different $\lambda$-values (Fig.~\ref{fig:4}a).
For the Standard GLASSO, $\lambda^* \sim 0.001$ and $\lambda \lesssim 0.1$ results in only one module in the corresponding network, providing no information about modular structure in the Covid-19 data. Excluding the edge-case peak for a disintegrated network with many singletons, $\lambda^* \sim 0.36$ for the Modular GLASSO resulting in 14 modules (Fig.~\ref{fig:4}b). The modules spread on the world map exhibit a geographic signal, with neighboring countries often belonging to the same module, as in Eastern Europe and parts of Central America, for example. China, however, forms its own module. In some cases, the connection between countries within the same module is less obvious, such as between the United States and the Iberian Peninsula, leaving it unclear whether a causal connection exists.

While the Standard GLASSO provides no information about modular structure in the global Covid-19 data, the Modular GLASSO unveils intriguing modular patterns. This situation resembles cases with high noise levels and many samples in the analysis of synthetic data when the Standard GLASSO retains many spurious relations, resulting in a dense, module-free network.

\paragraph{Gene co-expression data.} We analyze gene co-expression data obtained from the plant {\it Arabidopsis thaliana} under cold stress with included control samples (see Methods for details). We select the 1,000 genes with the highest variance across the 209 samples. Similar to the Covid-19 data, the correlations deviate significantly from what would be expected from pure noise (the Kolmogorov-Smirnov statistic is 0.39). In this case, however, the number of features exceeds the number of samples.

\begin{figure}[tb]
\centering
\includegraphics[scale=0.5]{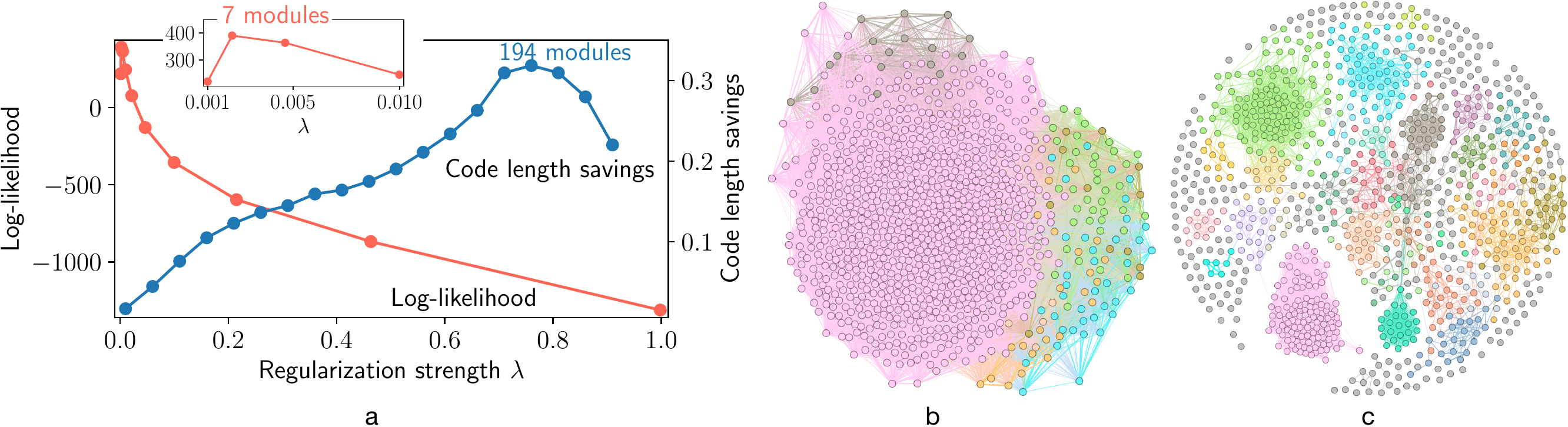}
\caption{\textbf{Application of Standard and Modular GLASSO to gene co-expression data}. The Standard GLASSO (log-likelihood) and Modular GLASSO (codelength savings) suggest vastly different regularization strengths also for the gene co-expression data (a). The Standard GLASSO's minimal regularization leads to a network with little modular structure (b). In contrast, Modular GLASSO disconnects nodes to maximize the modular structure during cross-validation, revealing more regularities in the underlying system (c).}
\label{fig:5}
\end{figure}

Cross-validating using the codelength savings to maximize the modular structure common to training and test data regularizes more. The Standard GLASSO applies minimal regularization ($\lambda^* \sim 0.002$) and finds seven modules in the data (Fig.~\ref{fig:5}a). In contrast, the Modular GLASSO suggests strong regularization ($\lambda^* \sim 0.76$) and disconnects nodes, resulting in distinct network representations of the data (Fig.~\ref{fig:5}bc).  The stronger regularization can reveal additional structure in the underlying data, offering valuable insights into the gene regulation patterns.

\section*{Discussion and conclusions}
Regularizing Gaussian graphical models is challenging due to the presence of noise and the complexity of high-dimensional data.
To tackle this issue, we introduce a regularization method that capitalizes on the modular structure inherent in the data. The Modular GLASSO outperforms standard regularization approaches by applying stronger regularization to retain only the connections that contribute significantly to the modular structure. In contrast, the Standard GLASSO, which maximizes the Gaussian log-likelihood, regularizes less when dealing with noisy data, retaining many noisy links and failing to detect modular structure. The differences between these two methods are crucial when analyzing real-world data sets. For example, when analyzing Covid-19 incidence data and gene co-expression data, the Modular GLASSO uncovered more modular structure in the data, providing deeper insights about the underlying system such as identifying groups of countries with similar epidemic patterns and gene clusters with similar functions. 

Constructing a network from correlational data requires many samples. When cross-validating modules, both the training and test networks must contain modular structure present in the complete data set. Two-fold splitting offers a reliable approach but requires relatively many samples, leading to suboptimal results with Modular GLASSO for low-noise data (Fig.~\ref{fig:2}) and relatively large spread (Fig.~\ref{fig:3}). A potential solution to this data-splitting issue is to eliminate the need for splitting altogether by employing Bayesian methods as an alternative to cross-validation. A Bayesian approach would make Modular GLASSO less data demanding.

Using codelength savings for model selection may result in selecting an overly sparse model when some modules have much larger link weights than others. In such cases, the codelength savings in the test network can be larger if the modules with smaller link weights are completely disintegrated into disconnected nodes. Partly washing out modular structure by excessive regularization in this way can, however, reveal potentially interesting structure through the remaining modules that can be difficult to discern with less regularization, as for the gene co-expression data (Fig.~\ref{fig:5}c). Since the disconnected nodes are weakly connected in the unregularized network, disconnecting them is supported in the data and in line with the module-based regularization.

In summary, we find that regularization based on modules effectively uncovers more structure in relational data sets. Because many downstream analysis tasks rely on identifying modular structure, including studying groups, communities, and clusters of observed features, many researchers may find it appealing and intuitive to base also their model selection criterion on modular structure. As we show, this approach is essential for detecting underlying modular structure in correlational data that would otherwise remain obscured by noise.

\section*{Methods}

\subsection*{Optimization and randomness in the map equation}
The relative codelength savings in the test network, $L^{test}(M^{\lambda,train})$, which depend on the regularization parameter $\lambda$, are stochastic for two reasons: the two-fold splitting in the cross-validation and the inherent randomness in the search algorithm Infomap optimizing the non-convex map equation objective function \cite{Calatayud}. To overcome this stochasticity, we perform two-fold splitting ten times and average the results. To calculate the AMI in Fig.~\ref{fig:2}, we also perform a sample average approximation with ten runs. For simple partition comparisons, we only look for two-level solutions with Infomap. 
When selecting the optimal $\lambda$ to calculate the AMI, we test a set of $\lambda \in \{0.01, 0.06, \ldots, 0.96\}$, and choose the $\lambda$ corresponding to the first maximum in the codelength savings. To avoid noise around zero at low $\lambda$ values, we use the additional condition that the codelength savings must exceed 0.01.

\subsection*{Analysis and real-world data}
We use the R packages {\tt GLASSO} and {\tt CVGLASSO} throughout the tests. To find the maximum log-likelihood, we increase or decrease the function parameters {\tt nlam} and {\tt lam.min.ratio} if necessary.

Gene co-expression data come from the Sequence Read Archive (SRA), where we identified all available RNA-Seq samples relating to cold stress in the leaf tissue of {\it Arabidopsis thaliana} ecotype Columbia-0. The selected data include both control and treated samples and were retrieved in April 2021. 
The data were quantified using salmon version 1.2.1 \cite{patro2017salmon} against the Araport 11 release of the {\it Arabidopsis thaliana} genome. Pre-processing and normalization were done in R using the variance stabilizing transform available in DESeq2 \cite{love2014moderated}.

To avoid constant values when analyzing the world Covid-19 data, we leave out countries belonging to the 10:th percentile with the lowest variance in daily incidence. 

All data are standardized before analysis, and have zero mean and unit standard deviation.

\subsection*{Modular GLASSO algorithm}
Algorithm 1 shows the pseudo code for Modular GLASSO. The code uses the R function {\tt GLASSO} to estimate the precision matrix $\Theta$ with a given value of the the regularization parameter $\lambda$. The code uses the Infomap algorithm to infer the modules $M$ in a network $\mathcal{G}(\Theta)$ and to calculate the codelength savings $l$ when a network is partitioned into a given modular structure.
\begin{algorithm}[hbt!]
\caption{Modular GLASSO}\label{alg:cap}
\begin{algorithmic}
\State Input: $X\in \mathbb{R}^{p\times q}$, $p$ - number of features (nodes), $q$ - number of samples
\State Output: $\mathcal{G}(\Theta^*)$, best model

\State $\Lambda \gets \{0.01, 0.06, \ldots, 0.96\}$ // set of $\lambda$ values to test
\State $\lambda^* \gets \mathrm{min}(\Lambda)$ // optimal $\lambda$
\State $l^* \gets 0$ // optimal codelength savings
\For{$\lambda \in \Lambda$} // These steps are repeated 10 times and averaged
        \State $X^{train} \gets \mathrm{sample}(X, \mathrm{fraction} = 0.5, \mathrm{axis} = 1)$
        \State $X^{test} \gets X \setminus X^{train}$
        \State $\mathcal{G}(\Theta^{train}) \gets \mathrm{GLASSO}(X^{train}, \lambda)$
        \State $M^{train} \gets \mathrm{Infomap}(\mathcal{G}(\Theta^{train}))$
        \State $\mathcal{G}(\Theta^{test}) \gets \mathrm{GLASSO}(X^{test}, \lambda)$
        \State $l \gets \mathrm{Infomap}(\mathcal{G}(\Theta^{test}), M^{train})$
    \If{$l > l^*$}
        \State $l^* \gets l$
        \State $\lambda^* \gets \lambda$
    \EndIf
\EndFor
\State $\mathcal{G}(\Theta^*) \gets \mathrm{GLASSO}(X, \lambda^*)$
\end{algorithmic}
\end{algorithm}

\end{document}